Evidence for Irradiation Triggered Nonuniform Defect Distribution In Multiharmonic Magnetic Susceptibility of Neutron Irradiated $YBa_2Cu_3O_{7-\delta}$.

V. Sandu[1], S. Popa[1], D. Di Gioacchino[2], P. Tripodi[3]

1-National Institute of Materials Physics, Bucharest-Magurele, POB-MG-7, R-76900 Romania

2-INFN-LNF, National Laboratory of Frascati, Via E. Fermi 40, 00044 Frascati, Italy

3-H.E.R.A., Hydrogen Energy Research Agency, Corso della Repubblica 448, 00049 Velletri, Italy

Corresponding author:

Dr. Viorel Sandu,

By 07.31.2005

Dept. of Physics, Kent State University, Kent, OH-44242, USA. Phone-330-672-5406, e-mail: sandu@physics.kent.edu.

After 08.01.2005

Dept. Low Temp. Phys. & Supercond., National Institute of Materials Physics, Bucharest-Magurele, POBox MG-7, code 077125, Romania, Phone +40-21-4930047/ext 152, e-mail: vsandu@infim.ro, viorelsandu51@yahoo.com

Evidence for Irradiation Triggered Nonuniform Defect Distribution In Multiharmonic Magnetic Susceptibility of Neutron Irradiated $YBa_2Cu_3O_{7-\delta.}$

V. Sandu[1], S. Popa1, D. Di Gioacchino[2], P. Tripodi[3]

1-National Institute of Materials Physics, Bucharest-Magurele, POB-MG-7, R-76900 Romania

2-INFN-LNF, National Laboratory of Frascati, Via E. Fermi 40, 00044 Frascati, Italy

3-H.E.R.A., Hydrogen Energy Research Agency, Corso della Repubblica 448, 00049 Velletri, Italy

## Abstract

Multiharmonic *ac*-magnetic susceptibility $\chi_1, \chi_2, \chi_3$, of neutron irradiated Li-doped $YBa_2Cu_3O_{7-x}$ has revealed a nonmonotonic dependence of all harmonics on the neutron fluence. The irradiation has a strongly depressive influence on the intergrain connection suggesting an increase of the effective thickness of the intergranular Josephson junction at a neutron fluence of $0.98 \times 10^{17}$ cm$^{-2}$. Less damaged are the intragrain properties. A spectacular enhancement of the superconducting intragranular properties reflected in the characteristics of all harmonics was observed at highest fluence $\Phi = 9.98 \times 10^{17}$ cm$^{-2}$. We assume that this effect results from the development of a space inhomogeneous distribution with alternating defectless and defect rich regions.

**Key words**: neutron irradiation, radiation damage, multiharmonic susceptibility, $YBa_2Cu_3O_7$

1. INTRODUCTION

Radiation effects in ceramic superconductors have applicative potential for current carrying superconducting cables as long as irradiation remain an attractive source of strong pinning and the depression of the superconducting properties is maintained at a satisfactory level.

The structure of these materials with high structural anisotropy, displaying multiple sublattices populated with ions having a large range of atomic masses (spanned from oxygen to barium), and combined with the presence of the free charge, makes them a complex challenge in the attempt to understand the connection between superconductivity and radiation damages. A sustained effort was focussed on this issue in the past decade and the data accumulated allowed a better picture of the processes concerning the superconducting state [1-8]. However, few attempts were dedicated to the creation of a reliable model regarding the evolution of the defect distribution and their influence on the superconducting properties. Except the model proposed by Kulikov et al. [9], which is focussed on the oxygen sublattices, we have no information concerning another one dedicated to cuprates. In the case of metallic systems, which have certain resemblance with cuprates, it was developed a more elaborate model consisting in a coupled nonlinear equations systems [10, 11]. This model has not only an uniform solution but also opens the possibility of the formation of an inhomogeneous distribution of defects which alternate with defectless regions at high neutron fluence. The latter effect can appear if the fraction of line defects exceeds a certain threshold [12]. The possibility of such space variation of the defect distribution can change the picture of a continuous suppression of the superconductivity due to neutron damages when the fluence is increased and leaves the possibility of a recovery of the superconducting properties exactly at high fluence.

In this paper, we present the results of multiharmonic magnetic susceptibility measurements on polycrystalline Li-doped $YBa_2Cu_3O_{7-x}$ which is consistent with the evolution of the defect distribution from a homogeneous one to the development of the space fluctuation of defect concentration.

## 2. EXPERIMENTAL

Superconducting Li-doped $YBa_2Cu_3O_{7-x}$ ceramics were obtained using solid state reaction from high purity reagents: $Y_2O_3$, $CuO$, $BaCO_3$ and small amount of lithium fluoride LiF (2 at. % Li). The reagents were calcined at 930 $^0$C for 20 hours, reground, pelletetized, and finally sintered in flowing oxygen for other 16 hours. The as obtained pellets were sealed in quartz ampoules and irradiated in standard aluminum blocks suspended in the center of the channel 36/6 of the reactor VVRS from the IFIN-HH Bucharest. The neutronic characteristics of this channel were determined [13] by absolute measurements and showed that more than 87 % of the neutrons have the energy below 1 MeV. The flux density is $2.13\times10^{13}$ neutrons/cm$^2$·sec. In our experiments we used the following fluences, $\Phi_1 = 0.98\times10^{17}$ cm$^{-2}$ and $\Phi_2 = 9.98\times10^{17}$ cm$^{-2}$. During the irradiation, the temperature inside the channel was below 40°C. Because the neutron irradiation causes also an activation of the material, we checked the samples activity immediately after irradiation. Activity was was between 1.8 $\mu$R and 80 $\mu$R, the samples were stored for seven days in the hot chamber and measured only after the activity decreased below the international standard for public area activity. It is to note that $YBa_2Cu_3O_{7-x}$ gives the lowest activity in this class of cuprates [14] after neutron irradiation.

The $ac$-susceptibilities, including the higher harmonics, were measured with a home made susceptometer [15] based on pick-up double coils surrounded by a driven coil. The sample was mounted on a sapphire holder inserted in one of the pick up coils. The temperature was measured with a platinum thermometer (PT100) in a good thermal contact with the samples. The whole assembly was cooled in ZFC, in a thermally controlled He gas flow cryostat provided with an 8 Tesla superconducting magnet. The measurements have been done on sweeping the temperature with a rate of 0.3 K/min up to a temperature greater than the zero field critical temperature of the samples (i.e. between 70 and 110 K). The *ac* driving magnetic field had an amplitude of $6 \times 10^{-4}$ T at a frequency of $f = 1070$ Hz. The *dc* magnetic field was in the range from 0 to 2 Tesla. The induced signal has been measured with a multi-harmonic EG&G lock-in amplifier. Both *ac* and *dc* fields were applied parallel to longest size of the sample.

## 3. RESULTS AND DISCUSSION

The effect of irradiation, as reflected in the fundamental harmonic af *ac*-susceptibility, in comparison with the unirradiated (virgin) samples is shown in Figure 1 for the fluences $\Phi_1 = 0.98 \times 10^{17}$ cm$^{-2}$ and $\Phi_2 = 9.98 \times 10^{17}$ cm$^{-2}$. The real part $\chi_1'$ of the unirradiated sample (Fig. 1 b), which accounts for the superconducting screening, shows three temperature regimes: *i)* a high temperature regime starting below the critical temperature $T_c = 91.5$ K, showing a rather steep transition due to the rapid divergence of the London penetration length close to $T_c$; *ii)* an intermediate regime, 80 K $< T <$ 89 K, where the dependence of $\chi_1'$ is linear in $T$ and is due to the temperature dependence of the effective magnetic permeability [16] ; and *iii)* a low $T$ regime, below 80 K, where the diamagnetic screening saturates. The strengthening of screening with decreasing temperature for 80 K $< T <$ 89 K is actually the result of a wide distribution of the critical temperatures in different regions of the pellets, mainly situated between the superconducting grains like a disordered matrix, which either displays a depressed superconductivity or ensures the Josephson coupling between grains. The imaginary part $\chi_1''$, which reflects the loss of energy during one ac-cycle displays a broad peak at $T_p = 84$ K, which is due to the intergranular contribution. $T_p$ is the temperature where the vortices reach the center of the sample. At higher temperatures, a shoulder in $\chi_1$" marks the intragranular temperature at which the Abrikosov vortices penetrate the grain and reach its center. As in the case of the real part $\chi_1$', the large peak is understood if one considers the sample as a collection of superconducting of Josephson coupled islands, each having a certain disorder level, which, with decreasing $T$, gives rise to larger and larger superconducting clusters which behave like a single grain denying the penetration of the field.

The irradiation with neutrons at $0.98 \times 10^{17}$ cm$^{-2}$ produces an extremely small shift of the onset of $\chi_1$' to lower temperatures but an important reduction of the diamagnetic screening. However, $\chi_1$' traks the temperature dependence of the virgin sample. Corespondingly, the imaginary part $\chi_1$" (Fig. 1a) shows a reduced intergranular peak with an increased breadth and a more singled out intragranular peak. At lower temperatures, $T < 80$ K, $\chi_1$" increases slightly. When fluence is increased ten times, at $9.98 \times 10^{17}$ cm$^{-2}$, $T_c$ decrease is hardly noticeable ($T_c = 91.1$ K), the intermediate regime disappears, and the diamagnetic screening is reduced 1.8 times in comparison with the virgin sample. Even more spectacular is the vanishing of the intergranular peak of $\chi_1$", but the increasing signal with decreasing temperature can be a hint to its shift toward lower temperatures. The intragranular peak is less sensitive to irradiation. It becomes well separated but much less slightly shifted than at previous fluence.

The effect of irradiation is more spectacular in the higher harmonics $\chi_2$ (Fig. 2) and $\chi_3$ (Fig. 3). The second harmonic, which in Bean model is supposed to be zero, displays for the virgin sample a complex structure caused by those effects which produce the asymmetry of the hysterezis loop (surface barriers, trapped fields, etc). In our case $\chi_2$ has the amplitude of the same order with the third harmonic. The effect of the irradiation is to wipe out this structure both in $\chi_2$' and $\chi_2$''. The large dip at 85 K, which is predicted by theoretical models [17-19], and the smaller one at 90 K, which we attribute to the intragrain contribution are not more visible in $\chi_2$' at the highest fluence, whereas the tiny, positive peak is enhanced and seems to become the tail of a large peak shifted at much lower temperatures (Fig. 2. b). $\chi_2$'' supports equivalent changes (Fig. 2.a). Namely, the small high temperature dip shifts from 90 K to 87 K while the large dip at 84.7 K reduces drastically its amplitude and shifts to low temperatures.

The third harmonics also displays reduced values after irradiation (Fig. 3). The small oscillations of $\chi_3$' (Fig. 3.b) close to irreversibility temperature shifts 2 K downward but becomes more emphasized. The large, negative dip at 86.6 K is broadened and shifted below 70 K. The imaginary part $\chi_3$'' (Fig. 3a) has a dip at 90 K which shifts at 88.8 K but reveals a high temperature shoulder after the irradiation with $\Phi_2$= 9.98×10$^{17}$ cm$^{-2}$ whereas the positive peak is reduced below 70 K so that the large dip of intergarnular origin is no more shown.

The Cole–Cole representation, $\chi_1$'' vs $\chi_1$' (Fig. 4a), shows that the expected double dome shape is dominated by the intergranular effects in the virgin sample and by the intragrain effects at the highest fluence in the temperature range investigated in our experiments (70-95 K). It is also relevant the important change in the Cole-Cole representation for $\chi_2$ (Fig. 4b) and $\chi_3$ (Fig. 4c), where the almost close circle in the case of virgin sample is reduced to a small arch.

As a common feature of all harmonics, the structures mirroring the intragranular contributions to $\chi$ are less modified by irradiation, while the intergranular signal are reduced and shifted with more than 10 K. This is a hint that, despite a homogeneous generation by irradiations, the damages accumulate mainly in the intergrain area.

A similar behavior occurs when a *dc*-magnetic field is superposed on the sample. Therefore, because it is accepted that the effect of a *dc*-magnetic field is extremely severe at the intergrain level, a comparison allows discriminating, in the higher harmonics, the intergranular and intragranular structures, hence, to evaluate their response to irradiation.

Figs. 5 show the evolution of the ac-susceptibility, fundamental and harmonics, when a dc-field of 0.1 T is applied. The similitude is striking for all harmonics, both for the real part (Fig 5, a, b, c) and imaginary part (Fig 5. d, e, f). As expected, the diamagnetic screening $\chi_1$' is reduced (Fig. 5 a), the intergranular peak is shifted to lower temperatures and whereas the intragranular peak in $\chi_1$" is singled out (Fig. 5 d). The dips in $\chi_2$' are cancelled (Fig. 5b), the breadth of $\chi_2$'' is increased and the amplitude is reduced (Fig. 5e). The structure at high temperature of $\chi_3$' is emphasized and the intragranular dip is shifted down. The negative dip in $\chi_3$" is broadened and the peak at 88 K is shifted to lower temperatures.

Because the aim of the irradiation studies is to improve the irreversibility and to extend the limits of the potential use of the superconductors, we have also investigated the behavior of the sample at usual dc-fields (0.5-2 T). The main observation is that the effects of irradiation are not more a monotonous function of fluence. Figs 6 and 7 show that in the presence of a dc-field the samples irradiated at the highest fluence display improved features even relative to the virgin sample. As can be seen from the real part $\chi_1$' (Fig. 6 a, 6 b, and 6 c), the virgin sample does not shows the saturation of the diamagnetic screening down to 70 K for 0.5 an 1 T and only at 2 T starts to saturate (Fig. 6 c) at a much lower value. The saturation is reached only in the irradiated samples. The transition to the saturation is larger in the sample irradiated at $\Phi_1 = 0.98 \times 10^{17}$ cm$^{-2}$ and much steeper for the sample irradiated at $\Phi_2 = 9.98 \times 10^{17}$ cm$^{-2}$. Moreover, for the latter the diamagnetic screening at high temperature is improved even in comparison with the virgin sample pointing to the self-organization of the defects, which increases the pinning inside the grains. At low temperatures however, $\chi_1$'($\Phi$) saturates at values higher than $\chi_1$'($\Phi = 0$). The imaginary part $\chi_1$" (Figs. 7 a, 7 b, and 7 c) shows a decrease in amplitude at $\Phi_1 = 0.98 \times 10^{17}$ cm$^{-2}$ for the intragranular peak and a outstanding increase at the highest fluence, except for $\mu_0 H = 2$ T (Fig. 7 c), where it becomes comparable to the peak of the virgin sample. It is noteworthy that this peak is much better separated at this high fluence. Concerning the intergranular peak, it seems to be suppressed, but the small increasing tendency at low temperatures hints toward a shift below 70 K.

The third harmonics (Figs. 8 and 9) display the same non-monotony regarding the dependence on the fluence. The structure details observed in $\chi_3'(\Phi=0)$ (Fig. 8 a, 8 b, and 8 c) disappear in $\chi_3'(\Phi_1)$ and get recovered in $\chi_3'(\Phi_2)$. Similarly, $\chi_3''$ does not display anymore the positive, low temperature values (Fig. 9a, b, c). The amplitude of the intragranular dip decreases at low fluence but becomes deeper at the highest fluence.

The resemblance between field and irradiation effects point to the intergrain and grain border regions as the weakest region where irradiations produces the most severe changes, as an applied magnetic field does. If the field dependence is well understood in terms of flux dynamics, the irradiation effects have to involve geometrical effects. Indeed, here we have a process of self-organization of the defects created by irradiation and controlled by the pre-irradiative microstructure.

Generally, the response of a superconductor at the neutron irradiation is the result of a complicated process that involves both the pre-iradiative microstructural defects and the neutron radiation damages. In this process, first, neutrons produce Frenkel defects, interstitials and vacancies, which diffuse fast toward sinks. There, they are absorbed depending on the specific absorption bias and strength of each sink [10, 11]. Interstitials or vacancies might also be thermally released back from sinks. The subsequent effect is the production of less mobile vacancy and interstitial loops by cascades, which shrink or develop by point defect absorption or release. Surface defects, voids and bubbles, follow a similar kinetic. The time or fluence dependence is strongly influenced by the anisotropy of the crystal structure and lead to a self –organization of nano- and microdefects. It was shown that the coupled equations describing the evolution of all types of defects have, besides the stationary, uniform solution, a space nonuniform one describing the onset of a nonuniform defect density. Specifficaly, with increasing the fluence, the homogeneous distribution becomes unstable and there is a tendency of accumulation of defects in localized space regions [12, 20]. Depending on the interplay of the in-cascade interstitial loop production, cascade collapse efficiency of vacancy loops, and the excess network bias, the space distribution of defects is either uniform or display sharp distributions around sinks. Computer simulations have shown that for large irradiation time, the stable modes which depicts the homogeneous distribution of the defects are replaced by a nonuniform mode, which, in a strongly anisotropic system, consists in a regular parallel system of stripes of defects alternating with "clean" material [12].

The evolution of the susceptibility with the fluence and its field dependence could be put in good agreement with this picture if we consider that the grain border is the sink with the highest strength and posses a large bias factor. Indeed, the grain border is characterised by a high density of defects, mainly disorder in the oxygen sublattice, which suppress the superconducting order parameter [21, 22]. At low fluence, there is an almost uniform distribution of the nanodefects within each grain. Therefore, even the diffusion of the point defects toward the stronger sink, which is the grain border is important, it remains an appreciable amount of defects within grain. They depress the superconducting order parameter, hence, decrease the inragranular peak, and the diamagnetic screening. Moreover, the point defect, which are absorbed by the border sinks, contribute to the increase of the effective thickness of the intergrain Josephson junction, and, subsequently, to a suppression of the intergranular critical current density $j_c$. The rapid decrease of $j_c$ is the results of its exponential dependence on the barrier thickness $d$ as $j_c \propto \exp(-d/\xi)$. The lower Josephson critical field decreases also as $H_{c1J} \propto d^{-1/2}$. Such effects were reported by Sen *et al.* [23] in proton irradiated polycrystalline Bi-2212 superconductor. Therefore, all samples measured at $\Phi_1 = 0.98 \times 10^{17}$ cm$^{-2}$ exhibit decreased intergranular Josephson coupling, reduced diamagnetic screening and depressed intragranula pinning due to the increase of the disorder within grains (mainly due to the oxygen atom displacements). The decrease of the second harmonic at $9.8 \times 10^{17}$ cm$^{-2}$ also mirrors a process of increase of the accumulation of defects, hence, pinning centers at the grain border preventing the effect of surface barriers (Figs. 2).

The spectacular recovery at the highest fluence $\Phi_2 = 9.98 \times 10^{17}$ cm$^{-2}$ is the effect of the self-organization of defects, which creates a sharp distribution of defect concentration in the sink areas, but also clean, defectless superconducting regions. If these clean areas have a size higher than the in-plane coherence length, they can provide good superconducting channels. On the other hand, the regions with high defect concentration and abrupt profile of the space variation of the defect distribution function provide the pinning force necessary to maintain high critical current densities, hence abrupt field profiles within grains. So, even in the presence of a *dc*-magnetic field, which is supposed to reduce the field gradient inside auperconducting grain, the intragrain characteristics (high temperature peak in $\chi_1$", and high temperature dip in $\chi_3$") are enhanced, even beyond the values displayed by the virgin samples.

In summary, the multiharmonic magnetic susceptibility investigation of YBa$_2$Cu$_3$O$_{7-x}$ ceramics irradiated with neutrons has shown an enhancement of the superconducting properties and of pinning strength at high fluence ($9.98 \times 10^{17}$ cm$^{-2}$) eventhough at low fluence ($0.98 \times 10^{17}$ cm$^{-2}$) these properties were severely reduced. We assume that this behavior arises from the development of self-organized microstructures of defects consisting in alternating clean regions and regions with a high concentration of defects. This effect occurs when the absorption bias of sinks exceeds the defect production bias.

ACKNOWLEDGEMENTS

The research was supported in part by the European Community under the TARI contract HPRI-CT-1999-00088, and by Romanian Ministry of Education and Science in the framework of the MATNANTECH program, project 260/2004. We would like to thank to Dr. Paolo Laurelli and Dr. Sergio Bertolucci for the special support.

REFERENCES

1.  F. M. Sauerzopf, H. P. Wiesinger, H. W. Weber, and G. W. Crabtree, Phys. Rev. B **51**, 6002 (1995).

2.  T. Yano, M. Pekala, A. Q. he, A. Wisniewski, and M. L. Jenkies, Physica C **247**, 55 (1995).

3.  F. Vasiliu, V. Sandu, P. Nita, S. Popa, E. Cimpoiasu, and M. C. Bunescu, Physica C **303**, 209 (1998).

4.  V. Sandu, S. Popa, J. Jaklovszky, and E. Cimpoiasu, J. Supercond. **11**, 251 (1998).

5.  F. M. Sauerzopf, Phys. Rev. B **57**, 10959 (1998).

6.  M. Zehetmayer, F. M. Sauerzopf, H. W. Weber, J. Karpinski, and M. Murakami, Physica C **383**, 232 (2002).

7.  K. Ueda, T. Kohara, M. Okuda, H. Kodaka, K. Myata, S. Watauchi, and I. Tanaka, Physica C **388**, 369, (2003).

8.  U. Topal, L. Dorosinskii, and H. Sozeri, Physica C **407**, 49 (2004).

9.  D. V. Kulikov, Yu. V. Trushin, F. M. Sauerzopf, M. Zehetmayer and H. W. Weber, Physica C **355**, 245 (2001).

10. R. Bulogh, B. L. Eyre, and K. Krishan, Proc. Roy. Soc. London **A 346**, 81 (1975).

11. N. M. Ghoniem and G. L. Kuchinski, Rad, Eff. **39**, 81 (1975).

12. N. M. Ghoniem, D. Walgraef, and S. J. Zinkle, J. Computer. Aid. Mater. Design **8**, 1 (2002).

13. C. Miron, C. Garlea, I. Garlea and V. Raducu, Rev. Roum. Phys. **31**, 813. (1968).

14. T. Shitamichi, M. Nakano, T. terai, M. Yamawaki, and T. Hoshiyaca C **392-396**, 254 (2003).

15. D. Di Gioacchino INFN-LNF European Facility TARI contract HPRI-CT-1999-00088.

16. K. H. Müller, Physica C **159**, 717 (1989)

17. T. Ishida and R. B. Goldfarb, Phys. Rev. B 41, 8937 (1990).

18. D. Di Gioacchino, F. Celani, P. Tripodi, A. M. Testa, and S. Pace, Phys. Rev. B **59**, 11539 (1999).

19. M. J. Quin and C. K. Ong Physica C **319**, 41 (1999).

20. D. Walgraef, J. Lauzeral, and N. M. Ghoniem, Phys. Rev. B **53**, 14782 (1996)

21. B. H. Moeckly, D. K. Lathrop, and R. A. Buhrman, Phys. Rev. B **47**, 400, (1993).

22. C. Betouras and R. Joynt, Physica C **250**, 256 (1995).

23. P. Sen, S. K. Bandyopadhyay, P. Barat, and M. Mukerjee-Sol. State Commun. 120, 201 (2001).

**Figure Captions**

**Figure 1**. Temperature dependence of the fundamental harmonic of ac-susceptibility for virgin sample and the samples irradiated at $0.98 \times 10^{17}$ $cm^{-2}$ and $9.98 \times 10^{17}$ $cm^{-2}$ . a) imaginary part $\chi_1$"; b) real part $\chi_1$'.

**Figure 2.** Temperature dependence of the second harmonic of ac-susceptibility for virgin sample and the samples irradiated $9.98 \times 10^{17}$ $cm^{-2}$ . a) imaginary part $\chi_2$"; b) real part $\chi_2$'.

**Figure 3**. Temperature dependence of the third harmonic of ac-susceptibility for virgin sample and the samples irradiated at $9.98 \times 10^{17}$ $cm^{-2}$. a) imaginary part $\chi_3$"; b) real part $\chi_3$'.

**Figure 4.** Cole-Cole representation for virgin sample and the samples irradiated at $0.98 \times 10^{17}$ $cm^{-2}$ and $9.98 \times 10^{17}$ $cm^{-2}$.a) First harmonic; b)second harmonic; c) third harmonic.

**Figure 5.** Temperature dependence of the first three harmonics of ac-susceptibility for virgin sample in zero field and for $\mu_0 H = 0.1$ T. a) $\chi_1$' ;b) $\chi_2$'; c) $\chi_3$'; d) $\chi_1$" ; e)$\chi_2$" f) $\chi_3$".

**Figure 6.** Temperature dependence of the fundamental harmonic of ac-susceptibility, real part $\chi_1$', for virgin sample and the samples irradiated at $0.98 \times 10^{17}$ $cm^{-2}$ and $9.98 \times 10^{17}$ $cm^{-2}$ at for different applied dc-fields. Left panel real part $\chi_1$'; a) $\mu_0 H = 0.5$ T; b) $\mu_0 H = 1.0$ T; c) $\mu_0 H = 2.0$ T.

**Figure 7.** Temperature dependence of the fundamental harmonic of ac-susceptibility, imaginary part$\chi_1$", for virgin sample and the samples irradiated at $0.98 \times 10^{17}$ $cm^{-2}$ and $9.98 \times 10^{17}$ $cm^{-2}$ at for different applied dc-fields: a) $\mu_0 H = 0.5$ T; b) $\mu_0 H = 1.0$ T; c) $\mu_0 H = 2.0$ T.

**Figure 8.** Temperature dependence of the real part of the third harmonic of ac-susceptibility $\chi_3$' for virgin sample and the samples irradiated at $0.98 \times 10^{17}$ $cm^{-2}$ and $9.98 \times 10^{17}$ $cm^{-2}$: a) $\mu_0 H = 0.5$ T; b) $\mu_0 H = 1.0$ T; c) $\mu_0 H = 2.0$ T;

**Figure 9.** Temperature dependence of the imaginary part of the third harmonic of ac-susceptibility $\chi_3$" for the virgin sample and the samples irradiated at $0.98 \times 10^{17}$ $cm^{-2}$ and $9.98 \times 10^{17}$ $cm^{-2}$: a) $\mu_0 H = 0.5$ T; b) $\mu_0 H = 1.0$ T; c) $\mu_0 H = 2.0$ T.

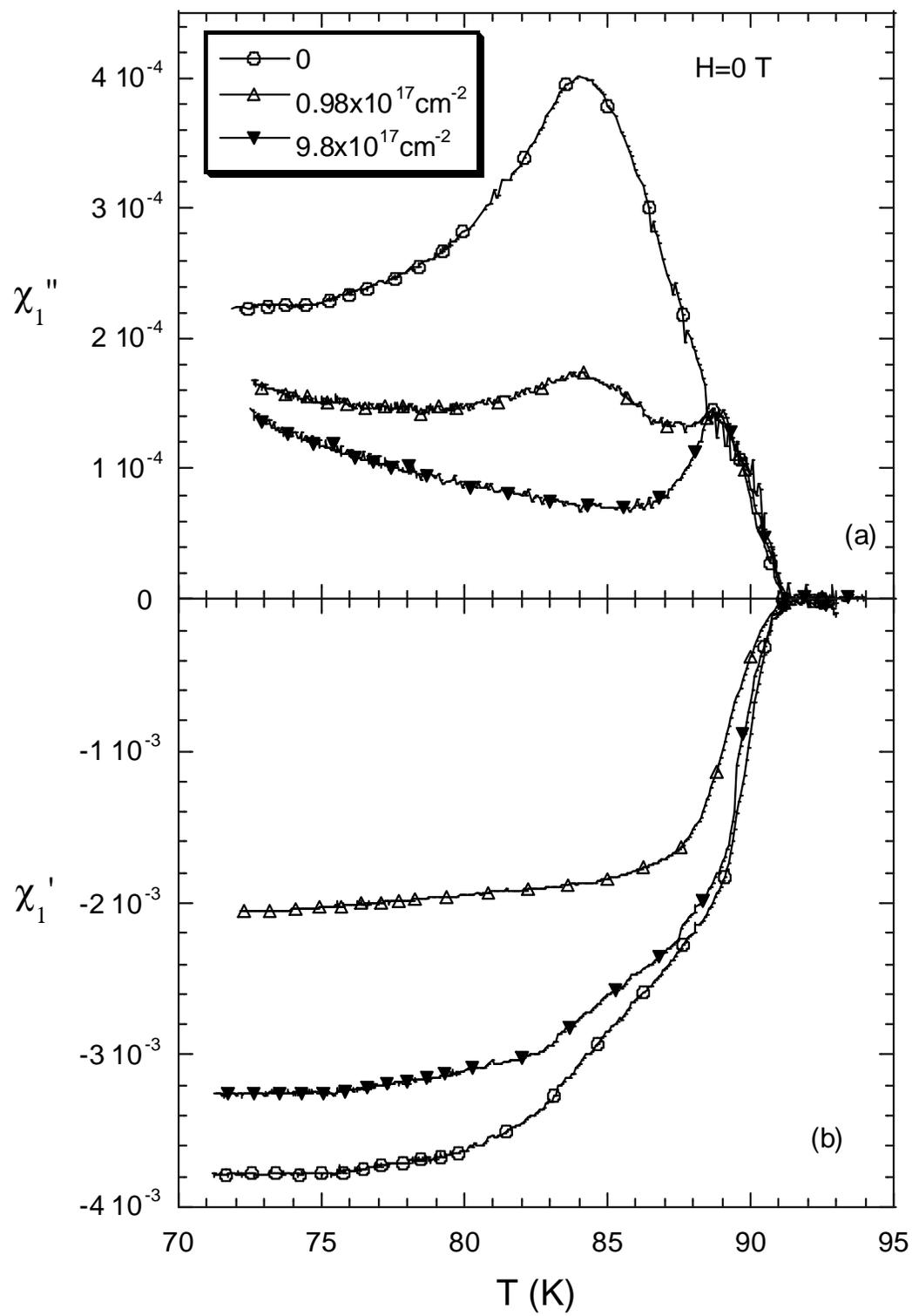

Fig.1

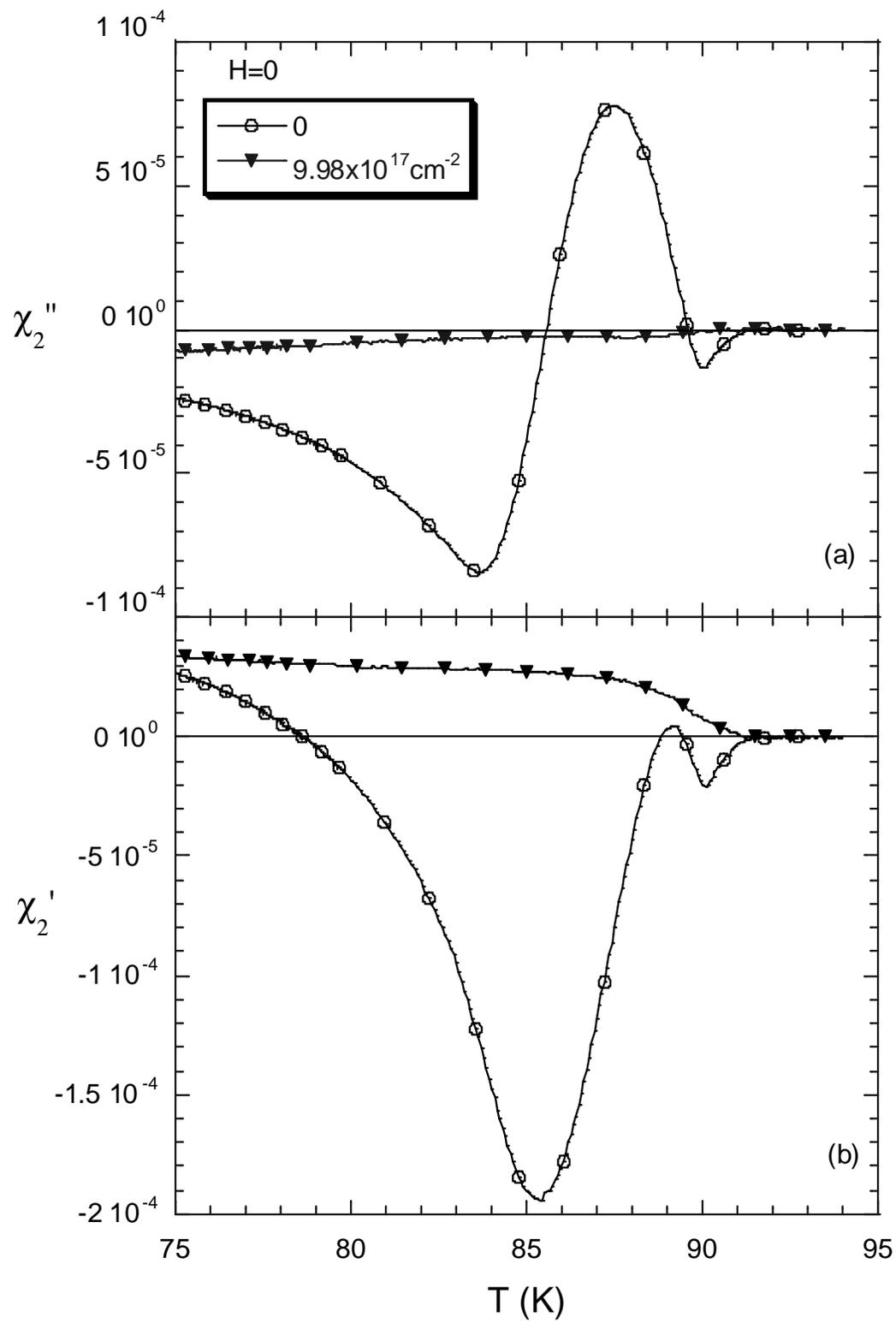

Fig. 2

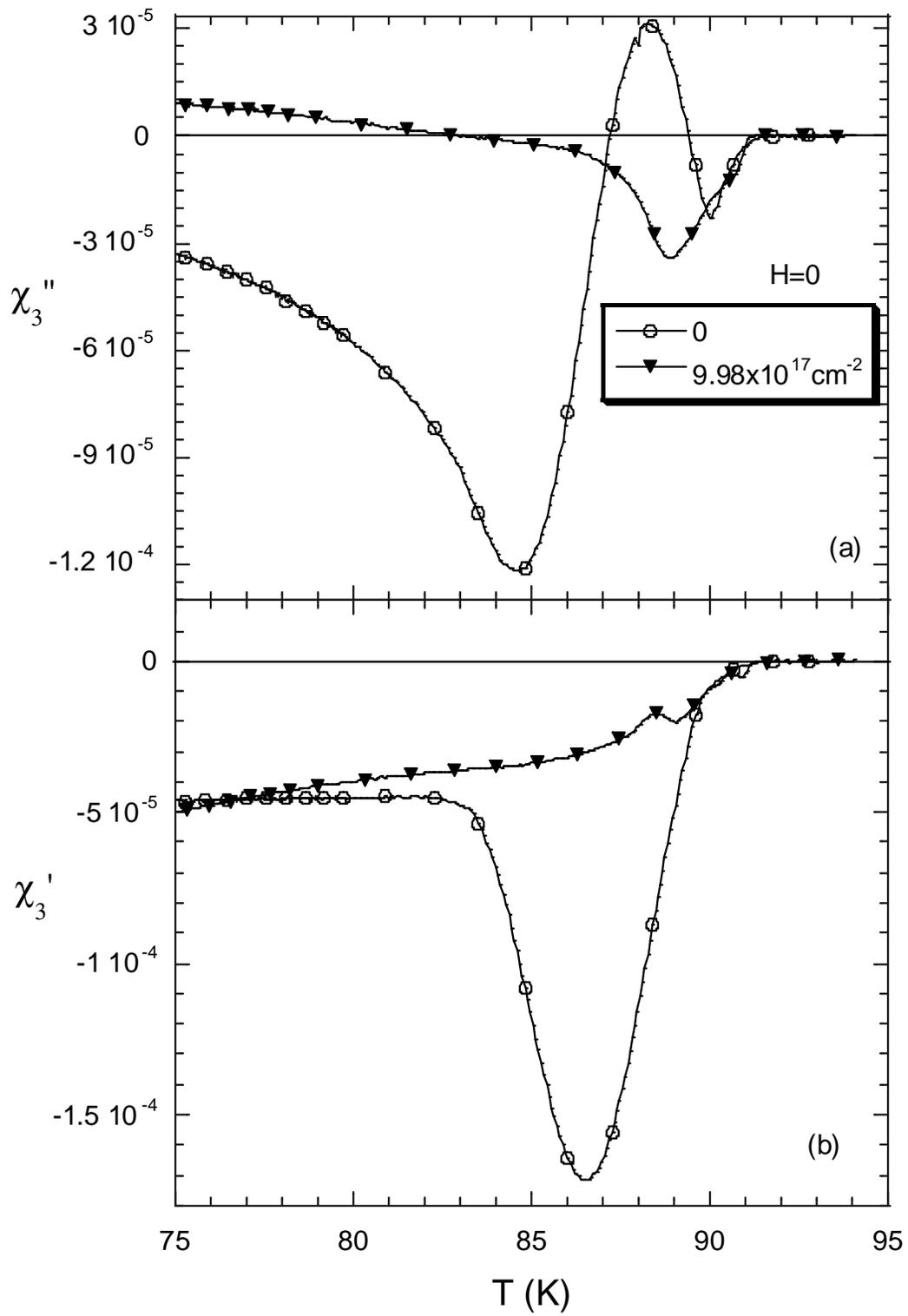

Fig. 3

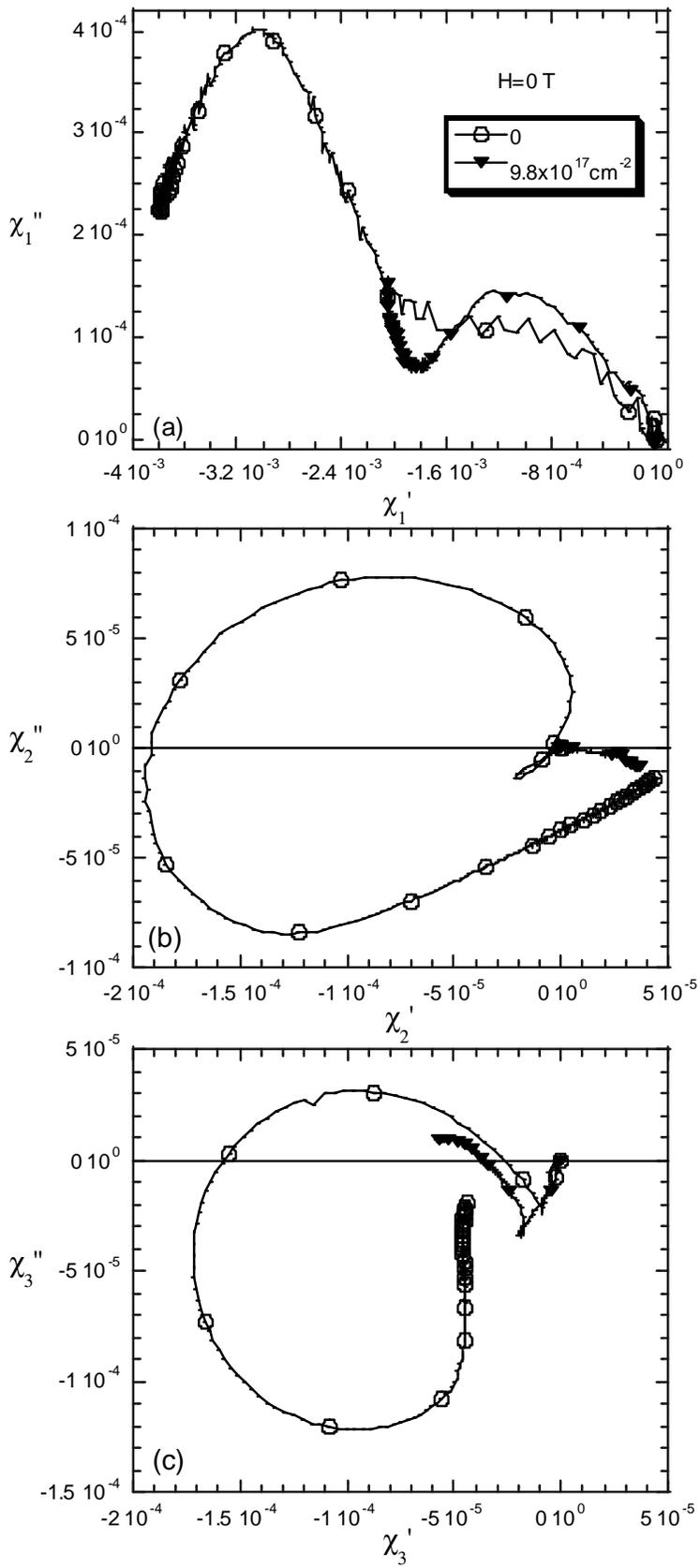

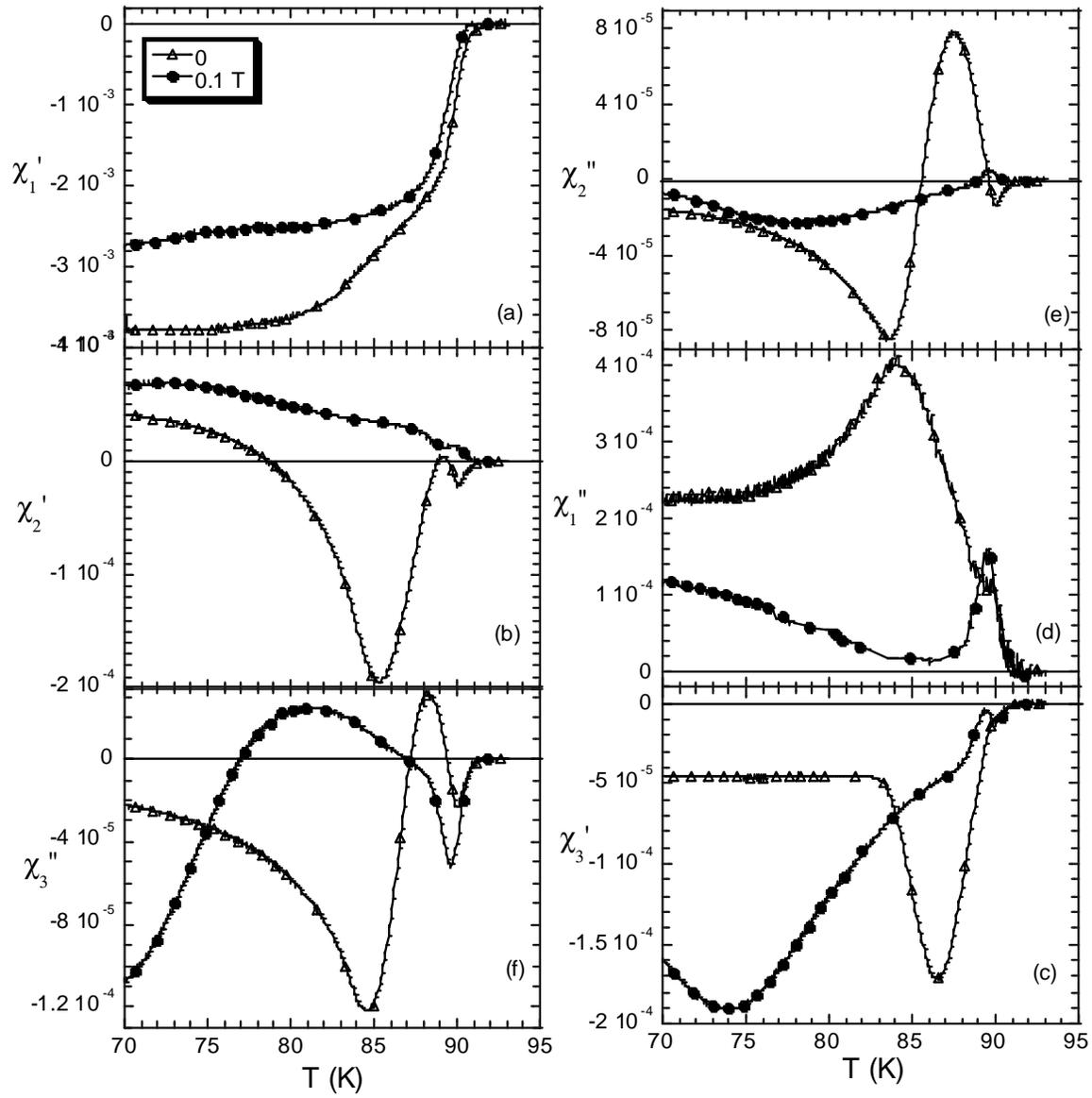

Fig. 5

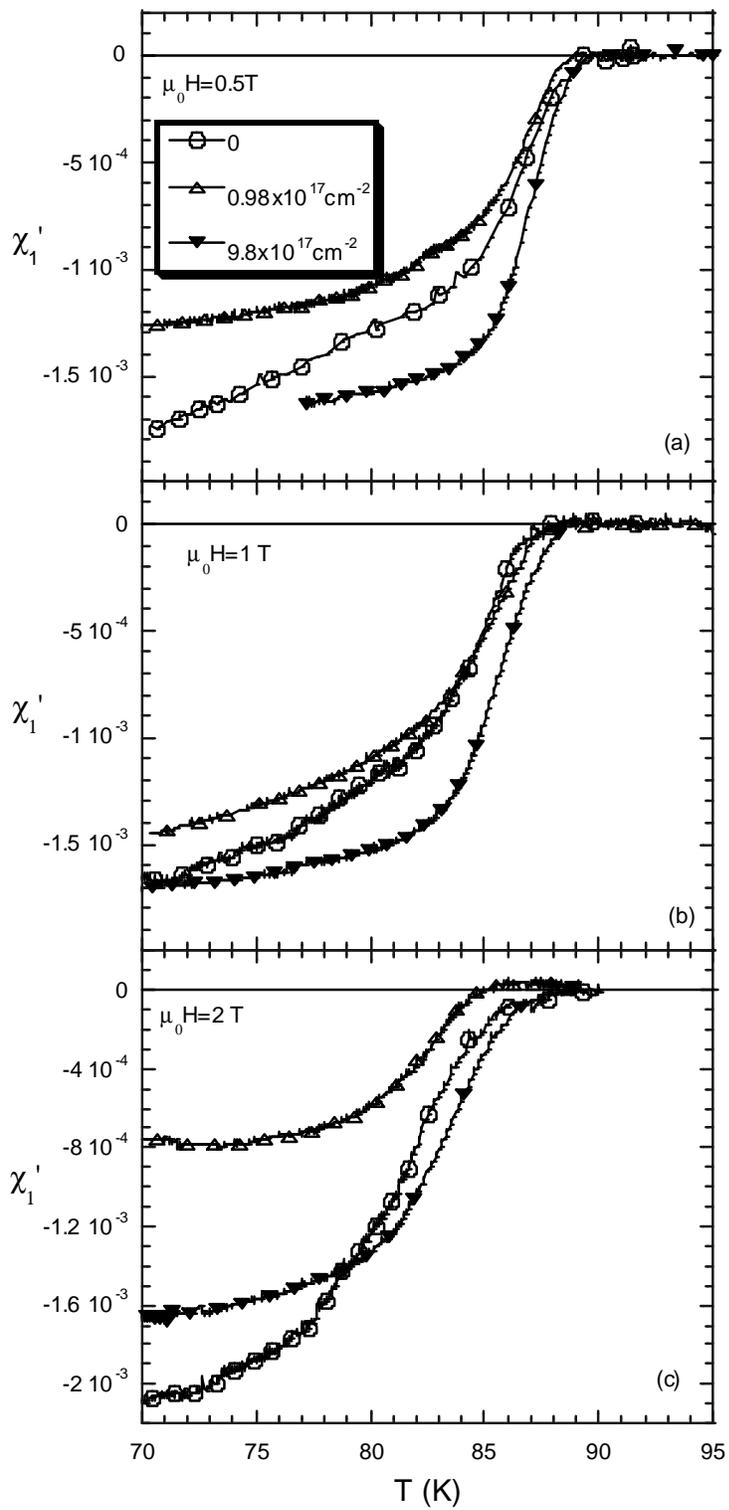

Fig. 6

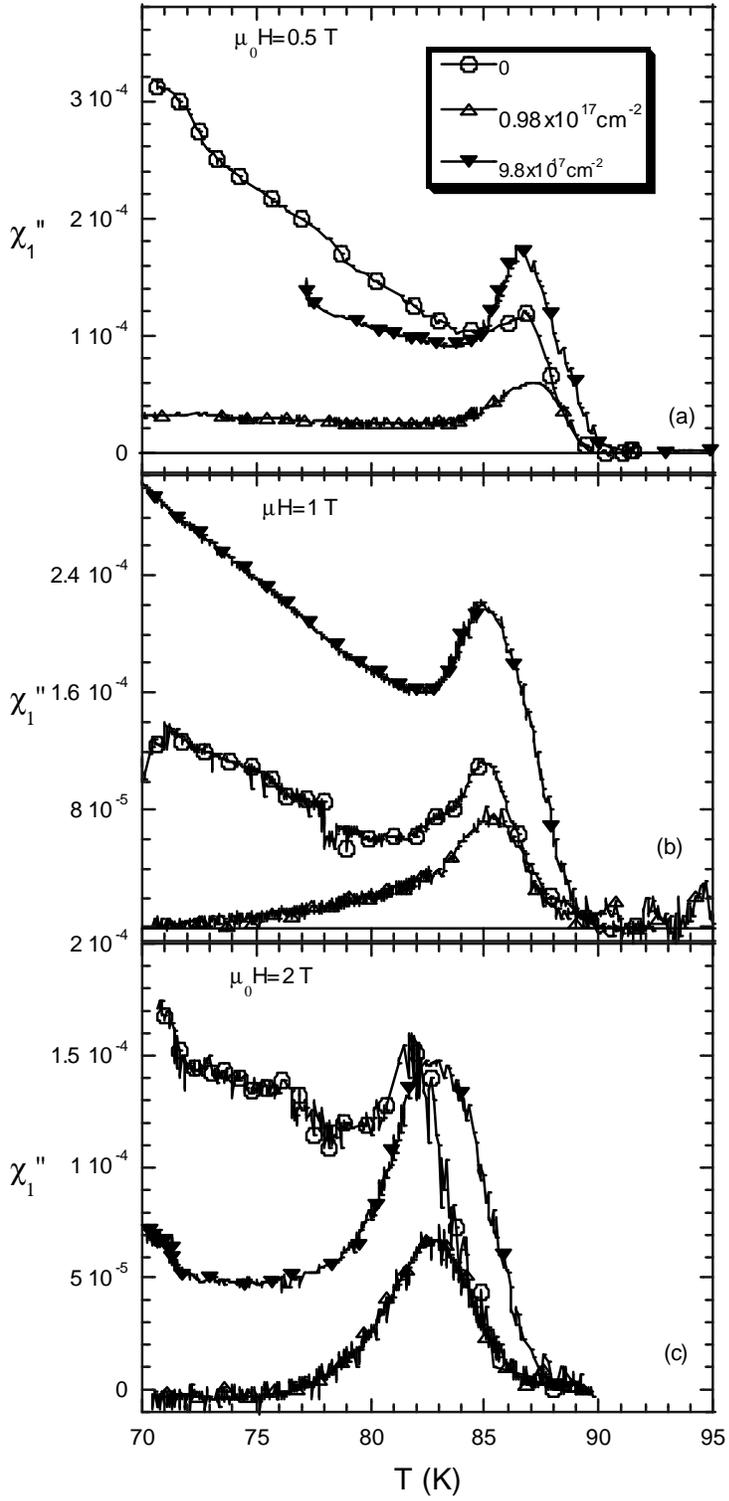

Fig. 7

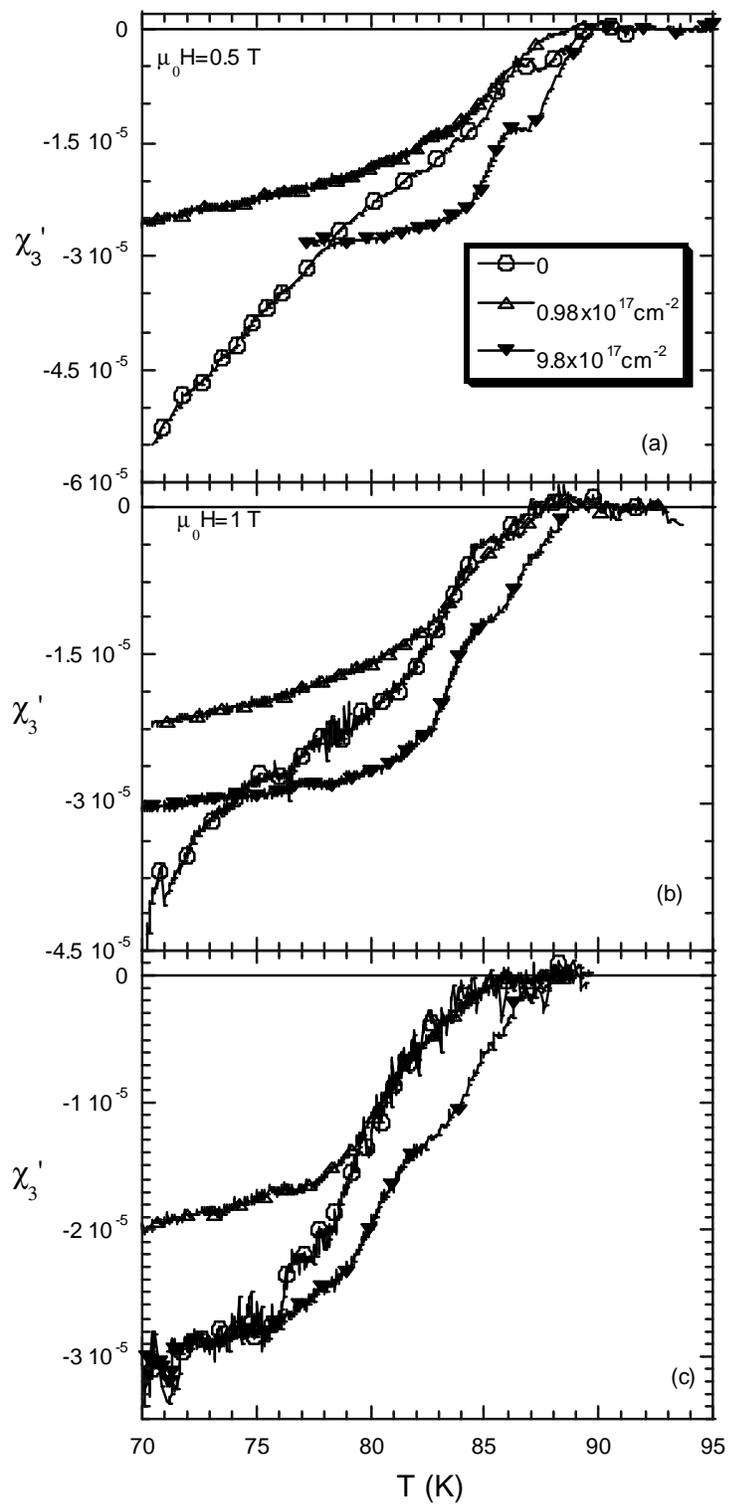

Fig. 8

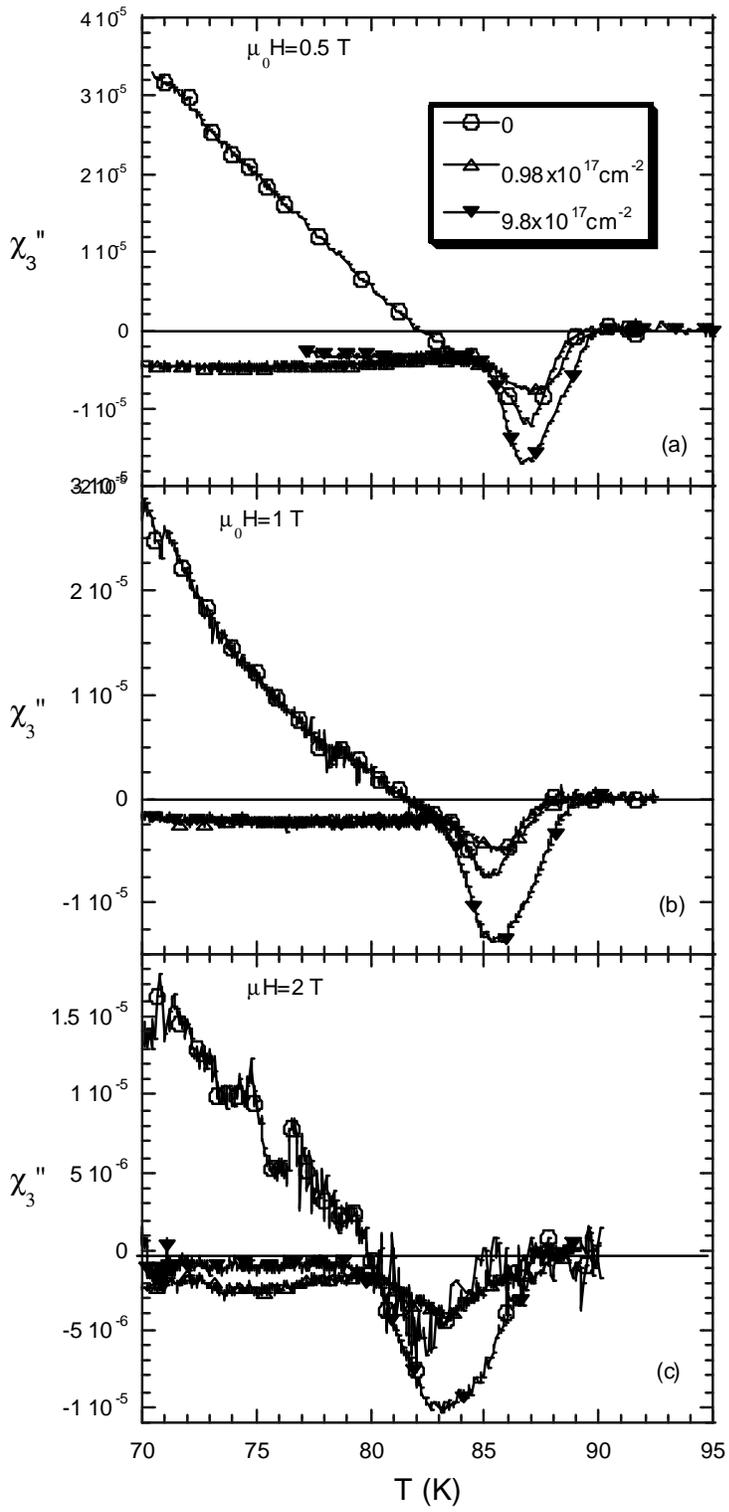

Fig. 9

Fig. 9